\documentclass[twocolumn]{aastex631}

\usepackage{graphicx}
\usepackage{color}
\usepackage{ulem}
\usepackage{float}
\usepackage{xcolor,soul}
\graphicspath{{fig/}}

\newcommand\hcnfour{HCN $J=4 \rightarrow 3$}

\newcommand\hcopfour{HCO$^+\ J=4 \rightarrow 3$}
\newcommand\coone{CO $J=1 \rightarrow 0$}
\newcommand\lx{$L_{\rm 0.5 - 2\,keV}^{\rm gas}$}
\newcommand\kms{km s$^{-1}$}
\newcommand\ldense{$L'_{\rm dense}$}
\newcommand\hcop{HCO$^+$}

\begin{document}

\title{Sub-kiloparsec scaling relations between hot gas, dense gas and star formation rate in five nearby star-forming galaxies}

\correspondingauthor{Junfeng Wang; Qing-Hua Tan}
\email{jfwang@xmu.edu.cn; qhtan@pmo.ac.cn}

\author{Chunyi Zhang}
\affiliation{Department of Astronomy, Xiamen University, 422 Siming South Road, Xiamen 361005, People’s Republic of China}

\author[0000-0003-4874-0369]{Junfeng Wang}
\affiliation{Department of Astronomy, Xiamen University, 422 Siming South Road, Xiamen 361005, People’s Republic of China}

\author{Qing-Hua Tan}
\affiliation{Purple Mountain Observatory, Chinese Academy of Sciences, 10 Yuanhua Road, Nanjing 210023, People’s Republic of China}

\author{Yu Gao}
\affiliation{Department of Astronomy, Xiamen University, 422 Siming South Road, Xiamen 361005, People’s Republic of China}

\author{Shuting Lin}
\affiliation{Department of Astronomy, Xiamen University, 422 Siming South Road, Xiamen 361005, People’s Republic of China}

\author{Xiaoyu Xu}
\affiliation{School of Astronomy and Space Science, Nanjing University, Nanjing 210093, China}
\affiliation{Key Laboratory of Modern Astronomy and Astrophysics, Nanjing University, Nanjing 210023, China}

\begin{abstract}
Based on the newly acquired dense gas observations from the JCMT MALATANG survey and X-ray data from \emph{Chandra}, we explore the correlation between hot gas and \hcnfour{}, \hcopfour{} emission for the first time at sub-kiloparsec scale of five nearby star-forming galaxies, namely M82, M83, IC 342, NGC 253, and NGC 6946. We find that both \hcnfour{} and \hcopfour{} line luminosity show a statistically significant correlation with the 0.5${-}$2 keV X-ray emission of the diffuse hot gas (\lx). The Bayesian regression analysis gives the best fit of ${\rm log}(L_{\rm 0.5-2\,keV}^{\rm gas} /{\rm erg\,s^{-1}})=2.39\,{\rm log}(L'_{\rm HCN(4-3)} /{\rm K\,km\,s^{-1}\,pc^{2}})+24.83$ and ${\rm log}(L_{\rm 0.5-2\,keV}^{\rm gas} /{\rm erg\,s^{-1}})=2.48\,{\rm log}(L'_{\rm HCO^{+}(4-3)} /{\rm K\,km\,s^{-1}\,pc^{2}})+23.84$, with dispersion of $\thicksim$0.69 dex and 0.54 dex, respectively. At the sub-kiloparsec scale, we find that the power-law index of the {\lx} ${-}$ star formation rate (SFR) relation is ${\rm log}(L_{\rm 0.5-2\,keV}^{\rm gas} /{\rm erg\,s^{-1}})=1.80\,{\rm log} ({\rm SFR} /M_\odot\,{\rm yr}^{-1})+39.16$, deviated from previous linear relations at global scale. This implies that the global property of hot gas significantly differs from individual resolved regions, which is influenced by the local physical conditions close to the sites of star formation. 


\end{abstract}


\section{Introduction}\label{sec:intro}

As building blocks of galaxies, understanding the birth, evolution, and eventual death of stars is fundamental to unraveling the key problems of galactic evolution. In the last few decades, it has become increasingly evident that the molecular gas, rather than atomic gas, is the raw material for star formation. \citet{1998Kennicutt} established a relationship between the global surface densities of total gas, traced by the H{\sc i} 21 cm line and rotational lines of CO, and the star-formation rate, with a power-law index of $n\approx 1.4$ ($\Sigma_{\rm SFR}\propto \Sigma^n_{\rm gas}$). In contrast, studies on dense gas (defined as molecular gas with a volume density $ n \gtrsim 10^4$ cm$^{-3}$) at (sub)millimeter bands indicate that the amount of dense molecular gas is tightly and linearly correlated with the SFR, from the dense molecular core of the Milky Way to high-redshift galaxies \citep{2004gaoa,2004gaob,2008Baan,2011wang&zhang&shi_416_L21,2012Garcia-Burillo&Usero_539_A8,2014ApJ...784L..31Z,2015Usero,2018ApJ...860..165T}. This suggests that the dense molecular gas is more closely associated with star formation. 

\begin{deluxetable*}{lccccccccc}
    \centering
    \tablecaption{The basic properties of sample galaxies \label{tab:table1}}
    \addtolength{\tabcolsep}{-0.2pt}
    \tablewidth{0pt}
    \tablehead{
    \colhead{Galaxy} & \colhead{R.A.} & \colhead{Decl.} & \colhead{Distance\tablenotemark{a}} & \colhead{Inclination} & \colhead{$L_{bol}$\tablenotemark{b}} & \colhead{SFR\tablenotemark{c}} & \colhead{Exp.Time\tablenotemark{d}} & \colhead{Hubble Type} & \colhead{Seq.} \\
    \colhead{} & \colhead{(J2000)} & \colhead{(J2000)} & \colhead{(Mpc)} & \colhead{(deg)} & \colhead{(erg s$^{\rm -1}$)} & \colhead{($M{_\odot}$ ${\rm yr}^{-1}$)} & \colhead{(ks)} & \colhead{} & \colhead{}
    }
    \startdata
    NGC 6946 & 20 34 52.3 & +60 09 13.2 & 4.7 & 30 & 2.88 $\times$ 10$^{40}$ & 7.1 & 228 & SAB(rs)cd & 1 \\
    IC 342 & 03 46 48.5 & +68 05 46.0 & 3.4 & 25 & 1.29 $\times$ 10$^{40}$ & 1.9 & 207 & SAB(rs)cd & 2 \\
    M82 & 09 55 52.4 & +69 40 46.9 & 3.5 & 66 & 2.75 $\times$ 10$^{41}$ & 11.2 & 406 & IO sp & 3 \\
    M83 & 13 37 00.9 & $-$29 51 56.0 & 4.8 & 27 & 9.12 $\times$ 10$^{40}$ & 9.1 & 328 & SAB(s)c & 4 \\
    NGC 253 & 00 47 33.1 & $-$25 17 19.7 & 3.5 & 76 & 1.23 $\times$ 10$^{41}$ & 4.2 & 234 & SAB(s)c & 5 \\
    \enddata
    \tablenotetext{a}{Reference for distances: (1) \citet{Poznanski2009ApJ}; (2) \citet{Wu2014AJ}; (3) \citet{Dalcanton2009ApJS}; (4) (5) \citet{Radburn-Smith2011ApJS}.}
    \tablenotetext{b}{Bolometric luminosities of hot gas for entire galaxies, corrected for both Galactic and intrinsic absorption (Sect~\ref{sec:hot gas and sfr}).}
    \tablenotetext{c}{Star formation rate. References as follows: (1) (2) \citet{Kennicutt2011PASP}; (3) (4) \citet{2003Cohen}; (5) \citet{Sanders2003AJ}.}
    \tablenotetext{d}{Total exposure time of \emph{Chandra}. The relevant ObsIDs are as follows: 10542, 10543, 10544, and 10545 for M82; 12994, 13202, and 13241 for M83; 7069, 22478, 22479, 22480, and 22482 for IC 342; 790, 3931, and 20343 for NGC 253; 1043, 4404, 4631, 4632, 4633, 13435, 17878, 19040, and 19887 for NGC 6946.}
    \vspace{-0.5cm}
    \end{deluxetable*}

However, the exact mechanism of how the physical and chemical parameters affect the process of star formation is still unclear. A threshold hypothesis argues that the SFR in a galaxy does not depend on the overall average gas density since stars only form when gas density is above $10^4$ cm$^{-3}$ \citep{2012Lada&Forbrich_745_190,2014Evans&Heiderman_782_114}. In contrast, there is alternative model that does not require a specific threshold density \citep{2015Elmegreen_814_L30,2018Elmegreen_854_16}. The model argues that the $L_{\rm IR}$ $-$ dense molecular line luminosity ($L'_{\rm dense}$) relation of \citet{2004gaoa,2004gaob} can be interpreted as a link between young stars and nearby collapsing gas with a free-fall time comparable to the age of stars \citep{2015Elmegreen_814_L30,2018Elmegreen_854_16}. Turbulence also plays an important role in star formation and can affect its efficiency by regulating the distribution of gas density \citep{2007Krumholz&Thompson_669_289}. Recent detailed observational studies  \citep{2015Usero,2016Bigiel,2019Jimenez_empire_880_127} have shown evidence for variations in the efficiency of star formation in dense gas across different regions of nearby galaxies, consistent with the turbulence-regulated postulate.

On the other hand, the late stages of stellar evolution and the death of stars are characterized by violent, high-energy processes, which lead to copious X-ray emission. Therefore, the X-ray band is naturally associated with the star formation related properties of galaxies \citep{1989ARA&A..27...87F,2006ARA&A..44..323F}. For example, the collective luminosity of high-mass X-ray binaries (HMXBs) shows a strong linear correlation with the global SFR \citep{2003Ranalli&Comastri_399_39,2005Hornschemeier&Heckman_129_86,2012Mineo1}, and the total stellar mass of a galaxy has a scaling relation with the X-ray emission of low-mass X-ray binaries (LMXBs) \citep{2004Gilfanov&Grimm_347_L57,2010Lehmer,2011Boroson,2012Zhang&Gilfanov_546_36}. 

The hot ionized gas at a temperature of $\thicksim$ 10$^6$$-$10$^7$ K is a source of diffuse X-ray emission \citep{Nardini2022hxga}. However, compared with the studies of dense gas, there has been little attention on the correlation between the hot gas and SFR at resolved scale. The local environment closely reflects the conditions of star formation and evolution, and in the nuclear regions of galaxies, the conditions can vary significantly from those of the galaxy as a whole. From this perspective, we study five nearby galaxies, M82, M83, IC 342, NGC 253, and NGC 6946, selected from the MALATANG (Mapping the dense molecular gas in the strongest star-forming galaxies) large program on the James Clerk Maxwell Telescope (JCMT).  MALATANG is the first systematic survey of the spatially resolved dense molecular gas traced by \hcnfour{} and \hcopfour{} emission. All these galaxies have been observed multiple times by the $Chandra$ X-ray Observatory, which offers superior spatial resolution. 

This work explores the correlation between diffuse X-ray emission (0.5-2 keV, hereafter \lx) and dense gas for the first time, focusing on the law of star formation at the sub-kiloparsec scale in the centers of galaxies. The structure of the letter is as follows: Section~\ref{sec:data} describes the observations and data reduction. In Section~\ref{sec:results and discussion}, the results of fitting the dense gas tracers and SFR against the diffuse X-ray emission are presented. The main findings are summarized in Section~\ref{sec:summary}. 

\begin{figure*}[htp]
    \centering 
    \includegraphics[scale=0.705]{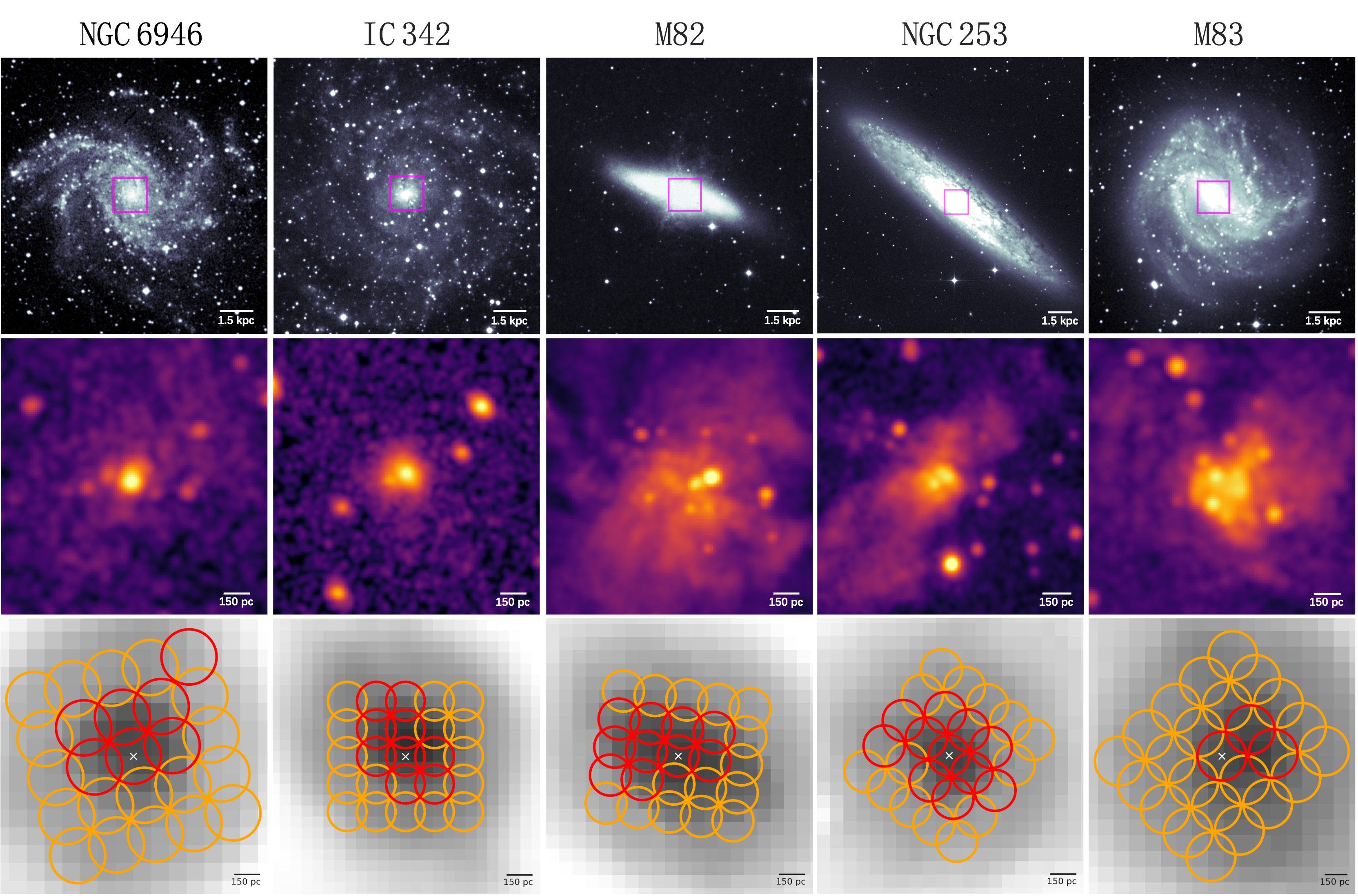}
    \vspace{0.1cm}
    \caption{ {\it Top row}: The \emph{R}-band images for NGC 6946 (6450\AA), IC 342 (6450\AA), M82 (6450\AA), NGC 253 (6400\AA), and M83 (6400\AA). Data downloaded from NASA/IPAC Extragalactic Database (NED). The purple box in each panel represents central\,$\thicksim$\,1.6 kpc region of each galaxy. {\it Middle row}: The {\em Chandra} broad band (0.5\,$-$\,7 keV) X-ray images of the central\,$\thicksim$\,1.6 kpc regions (purple box). {\it Bottom row}: The JCMT observing positions of the central $50\arcsec \times 50\arcsec$ regions of our sample galaxies overlaid on \emph{spitzer} MIPS 70 ${\mu m}$ emission on a logarithmic stretch. The 70 ${\mu m}$ images also show the central\,$\thicksim$\,1.6 kpc regions of our sample galaxies. White cross in each panel indicates the center of the galaxy. Colored circles denote positions observed in jiggle mode, with $14^{\prime \prime}$ diameters representing the FWHM of JCMT at about 350 GHz. Red circles in each panel indicate positions where HCN and HCO$^+$ are detected at S/N $\geqslant$ 3, and the derived properties of these positions are listed in Table~\ref{tab:table2}.}
    \vspace{0.4cm}
    \label{fig:f1a}
\end{figure*}

\section{Observations and data reduction} \label{sec:data}

\subsection{JCMT HCN (4$-$3) and HCO$^+$(4$-$3) Data}\label{subsec:jcmt}

For M82, IC 342 and NGC 253, the data of \hcnfour{} and \hcopfour{} are from \cite{2018ApJ...860..165T}. Because M83 and NGC 6946 were both observed by MALATANG \uppercase\expandafter{\romannumeral1} and MALATANG \uppercase\expandafter{\romannumeral2} survey from September 2015 to October 2021, we processed the MALATANG \uppercase\expandafter{\romannumeral1} and \uppercase\expandafter{\romannumeral2} data together. The observations of $J=4 \rightarrow 3$ lines of HCN and HCO$^+$ were obtained with the 16-receptor array receiver Heterodyne Array Receiver Program \citep[HARP;][]{2009MNRAS.399.1026B}. The full width at half maximum (FWHM) beamwidth of each receptor at 350 GHz is about $14^{\prime \prime}$ using a 3$\times$3 jiggle mode with grid spacing of $10^{\prime \prime}$ (see Figure~\ref{fig:f1a}). All the five sample galaxies were mapped in the central $2\arcmin \times 2\arcmin$ region. We measured the uncertainty in the absolute flux calibration using standard line calibrators and it was estimated to be about 10\%.

We reduced the raw data of M83 and NGC 6946 using the Starlink \citep[][]{2014ASPC..485..391C} software package ORAC-DR \citep[][]{2015MNRAS.453...73J} to obtain spectral data. It should be noted that MALATANG \uppercase\expandafter{\romannumeral1} observations used the position-switch (PSSW) mode for these two galaxies, which may have caused unstable baselines and high level noise. To enhance data quality, the beam-switch(BSW) mode was used during MALATANG \uppercase\expandafter{\romannumeral2} observations to obtain a steady baseline with low noise. After processing and inspecting the pipeline, we converted the spectra to the GILDAS/CLASS \footnote{\url{http://www.iram.fr/IRAMFR/GILDAS/}} format for further analysis. The spectral intensity units were converted from antenna temperature $T^\ast_{\rm A}$ to main beam temperature $T_{\rm mb}$ adopting a main beam efficiency of $\eta_{\rm mb}=0.64$ ($T_{\rm mb}\equiv T^\ast_{\rm A}/\eta_{\rm mb}$). We adopted the same criteria of \citet{2018ApJ...860..165T} (SNR $\geqslant$ 3) to identify detections of the emission lines. The uncertainty ($\sigma$) on the integrated line intensities was
\begin{equation}\label{eq1}
    \sigma=T_{\rm rms} \sqrt{\Delta v_{\rm line} \Delta v_{\rm res}} \sqrt{1+\Delta v_{\rm line}/\Delta v_{\rm base}},
    \end{equation}
where $T_{\rm rms}$ is the rms of the spectrum for a spectral velocity resolution of $\Delta v_{\rm res}$, $\Delta v_{\rm line}$ is the velocity range of the emission line, and $\Delta v_{\rm base}$ is the velocity width used to fit the baseline. For positions without significant detections, the 3$\sigma$ upper limits of the line integrated intensities are estimated. We derived the line luminosities $L^\prime_{\rm dense}$ for each position following \citet{1997solomon_478_144}:
\begin{eqnarray}
    L^\prime_{\rm dense}=&& 3.25\times10^7 \left(\frac{S\Delta v}{{\rm 1\ Jy\ km\ s^{-1}}}\right)\left(\frac{\nu_{\rm obs}}{{\rm 1\ GHz}}\right)^{-2} \nonumber\\ &&\times\left(\frac{D_{\rm L}}{{\rm 1\ Mpc}}\right)^2 \left(1+z\right)^{-3}\ {\rm K\ km\ s^{-1}\ pc^2},
    \end{eqnarray}
where $S\Delta v$ is the velocity-integrated flux density, $\nu_{\rm obs}$ is the observed line frequency, and $D_{\rm L}$ is the luminosity distance. For JCMT telescope at $\thicksim$ 350 GHz, we adopted the conversion factor of $S/T_{\rm mb}=15.6/\eta_{\rm mb}=24.4\ {\rm Jy\ K^{-1}}$. The luminosities where the $J=4 \rightarrow 3$ lines of HCN and HCO$^+$ $\geqslant$ 3$\sigma$ detection are listed in Table \ref{tab:table2}. The more detailed description of MALATANG survey and reduction strategy are given in Zhang et al. \citep[in prep., also in][]{2018ApJ...860..165T}. 


\begin{deluxetable*}{lrcccccc}[htp]
\centering
\vspace{-0.165cm}
\tablecaption{Properties for positions with significant ($\geqslant$\,3$\sigma$) HCN or HCO$^+$ detections of the five galaxies}
\addtolength{\tabcolsep}{0.5pt}
\tablewidth{0pt}
\tablehead{
\colhead{Galaxies} & \colhead{Offset\tablenotemark{a}} & \colhead{log$L'_{\rm HCN(4-3)}$\tablenotemark{b}} & \colhead{log$L'_{\rm HCO^+(4-3)}$} & \colhead{log$L'_{\rm CO(1-0)}$} & \colhead{log$L_{\rm IR}$} & \colhead{log\lx \tablenotemark{c}}& \colhead{c-stat/dof} \\ 
\colhead{} & \colhead{(arcsec)} & \colhead{(K \kms \ pc$^2$)} & \colhead{(K \kms \ pc$^2$)} & \colhead{(K \kms \ pc$^2$)} & \colhead{($L_{\odot}$)} & \colhead{(erg s$^{\rm -1}$)} & \colhead{} 
}

\startdata
NGC 6946 & (10,0) & 5.1 $\pm$ 0.07 & $\textless$ 5.0 & 7.2 $\pm$ 0.01 & 8.6 $\pm$ 0.02 & 36.42 $\pm$ 0.57 & 50/50 \\
                & (0,0) & 5.4 $\pm$ 0.04 & 5.4 $\pm$ 0.03 & 7.4 $\pm$ 0.01 & 9.3 $\pm$ 0.02 & 36.89 $\pm$ 0.09 & 96/72 \\
                & (-10,0) & 5.0 $\pm$ 0.14 & 5.2 $\pm$ 0.12 & 7.3 $\pm$ 0.01 & 9.1 $\pm$ 0.02 & 36.36 $\pm$ 0.18 & 70.49/42 \\
                & (10,10) & 4.9 $\pm$ 0.11 & 5.0 $\pm$ 0.13 & 7.2 $\pm$ 0.01 & 8.4 $\pm$ 0.02 & 35.70 $\pm$ 0.47 & 39/26 \\
                & (0,10) & 5.0 $\pm$ 0.08 & 5.3 $\pm$ 0.06 & 7.3 $\pm$ 0.01 & 8.9 $\pm$ 0.02 & 36.47 $\pm$ 0.08 & 94.97/56\\
                & (-10,10) & $\textless$ 5.0 & 5.2 $\pm$ 0.06 & 7.3 $\pm$ 0.01 & 8.8 $\pm$ 0.02 & 36.66 $\pm$ 0.35 & 43.49/31 \\
                & (-20,20) & 4.9 $\pm$ 0.11 & $\textless$ 5.1 & 7.0 $\pm$ 0.01 & 8.2 $\pm$ 0.03 & 36.30 $\pm$ 0.36 & 31.25/32 \\
IC 342 & (0,-10) & $\textless$ 4.7 & 5.1 $\pm$ 0.07 & 7.1 $\pm$ 0.01 & 8.7 $\pm$ 0.02 & 36.40 $\pm$ 0.18 & 37.58/34  \\
                  & (-10,-10) & 4.8 $\pm$ 0.14 & 4.8 $\pm$ 0.07 & 7.0 $\pm$ 0.01 & 8.5 $\pm$ 0.03 & 35.97 $\pm$ 0.24 & 14.16/20 \\
                  & (10,0) & 5.1 $\pm$ 0.07 & 4.8 $\pm$ 0.14 & 7.0 $\pm$ 0.01 & 8.9 $\pm$ 0.03 & 36.61 $\pm$ 0.44 & 33.42/47 \\
                  & (0,0) & 5.3 $\pm$ 0.04 & 5.4 $\pm$ 0.03 & 7.2 $\pm$ 0.01 & 9.3 $\pm$ 0.03 & 36.91 $\pm$ 0.09 & 28.31/23 \\
                  & (-10,0) & 5.0 $\pm$ 0.10 & 5.1 $\pm$ 0.07 & 7.0 $\pm$ 0.01 & 8.9 $\pm$ 0.02 & 36.14 $\pm$ 0.19 & 24.25/27 \\
                  & (10,10) & 5.0 $\pm$ 0.08 & 5.1 $\pm$ 0.06 & 7.1 $\pm$ 0.01 & 8.7 $\pm$ 0.03 & 36.02 $\pm$ 0.32 & 30.53/30 \\
                  & (0,10) & 5.1 $\pm$ 0.03 & 5.2 $\pm$ 0.05 & 7.1 $\pm$ 0.01 & 9.0 $\pm$ 0.03 & 36.56 $\pm$ 0.25 & 36.66/36 \\
                  & (10,20) & $\textless$ 4.7 & 4.8 $\pm$ 0.14 & 7.0 $\pm$ 0.01 & 8.0 $\pm$ 0.04 & 35.29 $\pm$ 0.33 & 7.78/6 \\
                  & (0,20) & $\textless$ 4.8 & 4.8 $\pm$ 0.12 & 7.0 $\pm$ 0.01 & 8.1 $\pm$ 0.03 & 35.12 $\pm$ 0.56 & 7.56/6 \\
M82 & (20,-10) & $\textless$ 5.1 & 5.8 $\pm$ 0.05 & 7.6 $\pm$ 0.01 & 9.5 $\pm$ 0.03 & 38.48 $\pm$ 0.03 & 103.51/87 \\
            & (10,-10) & 5.5 $\pm$ 0.06 & 5.9 $\pm$ 0.06 & 7.6 $\pm$ 0.01 & 9.5 $\pm$ 0.03 & 38.48 $\pm$ 0.10 & 103.51/87 \\
            & (20,0) & 5.4 $\pm$ 0.06 & 6.0 $\pm$ 0.03 & 7.6 $\pm$ 0.01 & 9.5 $\pm$ 0.03 & 38.48 $\pm$ 0.08 & 103.51/87 \\
            & (10,0) & 5.8 $\pm$ 0.02 & 6.3 $\pm$ 0.01 & 7.7 $\pm$ 0.01 & 9.9 $\pm$ 0.03 & 39.49 $\pm$ 0.04 & 115.01/87 \\
            & (0,0) & 5.9 $\pm$ 0.04 & 6.3 $\pm$ 0.01 & 7.7 $\pm$ 0.01 & 10.0 $\pm$ 0.03 & 39.67 $\pm$ 0.02 & 122.41/90 \\
            & (-10,0) & 5.8 $\pm$ 0.02 & 6.3 $\pm$ 0.02 & 7.7 $\pm$ 0.01 & 9.9 $\pm$ 0.03 & 39.12 $\pm$ 0.15 & 140.1/88 \\
            & (20,10) & $\textless$ 5.0 & 5.4 $\pm$ 0.12 & 7.6 $\pm$ 0.01 & 9.5 $\pm$ 0.03 & 38.6 $\pm$ 0.08 & 89.56/86 \\
            & (10,10) & 5.4 $\pm$ 0.04 & 5.9 $\pm$ 0.05 & 7.7 $\pm$ 0.01 & 9.8 $\pm$ 0.03 & 39.08 $\pm$ 0.09 & 104.15/80 \\
            & (0,10) & 5.4 $\pm$ 0.08 & 5.8 $\pm$ 0.06 & 7.6 $\pm$ 0.01 & 10.0 $\pm$ 0.03 & 39.14 $\pm$ 0.19 & 140/86 \\
            & (-10,10) & 5.5 $\pm$ 0.07 & 5.9 $\pm$ 0.03 & 7.5 $\pm$ 0.01 & 10.0 $\pm$ 0.03 & 38.58 $\pm$ 0.01 & 128.97/88 \\
NGC 253 & (20,-10) & 5.5 $\pm$ 0.04 & 5.5 $\pm$ 0.10 & 7.4 $\pm$ 0.01 & 8.6 $\pm$ 0.03 & 37.63 $\pm$ 0.02 & 75/54 \\
                  & (0,-10) & 5.9 $\pm$ 0.02 & 6.2 $\pm$ 0.02 & 7.5 $\pm$ 0.01 & 9.3 $\pm$ 0.03 & 39.95 $\pm$ 0.02 & 127.11/90 \\
                  & (-10,-10) & 5.4 $\pm$ 0.09 & 5.8 $\pm$ 0.04 & 7.4 $\pm$ 0.01 & 8.8 $\pm$ 0.03 & 38.60 $\pm$ 0.28 & 107.13/81 \\
                  & (20,0) & 5.9 $\pm$ 0.02 & 5.9 $\pm$ 0.04 & 7.8 $\pm$ 0.01 & 9.2 $\pm$ 0.03 & 38.32 $\pm$ 0.17 & 99.19/77 \\
                  & (10,0) & 6.4 $\pm$ 0.01 & 6.4 $\pm$ 0.01 & 7.9 $\pm$ 0.01 & 9.9 $\pm$ 0.03 & 39.59 $\pm$ 0.09 & 109.78/88 \\
                  & (0,0) & 6.6 $\pm$ 0.01 & 6.8 $\pm$ 0.01 & 7.9 $\pm$ 0.01 & 10.2 $\pm$ 0.03 & 40.44 $\pm$ 0.06 & 88.12/86 \\
                  & (-10,0) & 6.2 $\pm$ 0.01 & 6.4 $\pm$ 0.01 & 7.7 $\pm$ 0.01 & 9.8 $\pm$ 0.03 & 39.12 $\pm$ 0.07 & 85.55/79 \\
                  & (-20,0) & 5.8 $\pm$ 0.04 & 5.7 $\pm$ 0.06 & 7.6 $\pm$ 0.01 & 8.8 $\pm$ 0.03 & 37.98 $\pm$ 0.31 & 25.28/17 \\
                  & (10,10) & 5.7 $\pm$ 0.07 & 5.7 $\pm$ 0.06 & 7.7 $\pm$ 0.01 & 9.7 $\pm$ 0.03 & 38.84 $\pm$ 0.08 & 92.65/84 \\
                  & (0,10) & 6.1 $\pm$ 0.02 & 5.8 $\pm$ 0.04 & 7.7 $\pm$ 0.01 & 10.1 $\pm$ 0.03 & 39.59 $\pm$ 0.09 & 85.7/83 \\
                  & (-10,10) & 5.8 $\pm$ 0.02 & 5.8 $\pm$ 0.03 & 7.5 $\pm$ 0.01 & 9.7 $\pm$ 0.03 & 38.43 $\pm$ 0.08 & 54.7/55 \\ 
M83 & (0,0) & 5.4 $\pm$ 0.08 & 5.7 $\pm$ 0.06 & 7.3 $\pm$ 0.01 & 9.6 $\pm$ 0.03 & 38.98 $\pm$ 0.14 & 85.12/80 \\
                  & (0,10) & 5.4 $\pm$ 0.06 & 5.3 $\pm$ 0.15 & 7.2 $\pm$ 0.01 & 8.9 $\pm$ 0.03 & 38.21 $\pm$ 0.18 & 97/67 \\
                  & (-10,10) & 5.2 $\pm$ 0.11 & 5.5 $\pm$ 0.10 & 7.2 $\pm$ 0.01 & 9.0 $\pm$ 0.03 & 38.01 $\pm$ 0.04 & 51.1/49 \\
\enddata
\tablecomments{Except for the diffuse X-ray luminosities, all errors are estimated statistically from the measurements.}
\tablenotetext{a}{Offset along the major and minor axes of the galaxies, respectively.}
\tablenotetext{b}{For HCN and HCO$^+$ emission, the undetected positions are reported a 3$\sigma$ upper limit.}
\tablenotetext{c}{Diffuse X-ray luminosity corrected for Galactic and intrinsic absorption, and errors are quoted at 1-sigma confidence level.}
\vspace{-0.6cm}
\label{tab:table2}
\end{deluxetable*}

\subsection{X-ray Data}\label{sec:chandra}

All X-ray data are obtained from the $Chandra$ Data Archive, with a full list summarized in Table~\ref{tab:table1}. We first used the \emph{Chandra$\_$repro} task to reprocess the data with CIAO v.4.13 and CALDB v.4.9.6. Figure~\ref{fig:f1a} shows the \emph{R}-band and the X-ray maps based on the $Chandra$ data for the five galaxies. To examine the diffuse emission from the hot interstellar medium, we used the wavelet-based source detection algorithm \emph{wavdetect} \citep{Freeman2002ApJS} to search for discrete sources on the scales of 1, $\sqrt{2}$, 2, 2$\sqrt{2}$, 4, and 8 pixels ($0^{\prime \prime}.492$/pixel) and in the soft (0.5 ${-}$ 2.0 keV), hard (2.0 ${-}$ 8.0 keV) and total (0.5 ${-}$ 8.0 keV) energy bands. For most point sources covered by our JCMT observations of the five galaxies, excluding the aperture of 90\% energy-encircled fraction of the point spread function (PSF) was sufficient. For a few bright point sources, we manually increased the size of the elliptical mask to remove visible ring-like features due to the PSF wing to minimize the contamination from the PSF spillover.

The diffuse X-ray emission spectra of the regions matching the FWHM $\thicksim$ 14$\arcsec$ of JCMT beam were extracted using the CIAO task \emph{specextract}, and the background were chosen from the source free regions within the same field. We combined the multiple observed spectra of each position by the \emph{combine$\_$spectra} task with appropriate grouping. The XSPEC v.12.11.0 \citep[][]{1996ASPC..101...17A} software was used to fit the spectra. The models included an absorbed thermal plasma \citep[APEC;][]{2001ApJ...556L..91S}, and when necessary, Gaussian emission-line components. The best-fit diffuse X-ray luminosities corrected for both Galactic and intrinsic absorption were shown in Table~\ref{tab:table2}.

\subsection{Ancillary Data}\label{sec:ancillary}

In this work, we also utilize the \coone{} data from the Nobeyama CO Atlas of Nearby Spiral Galaxies  \citep[][]{2000PASJ...52..785S,2007PASJ...59..117K}, and infrared archival imaging data from \emph{Spitzer} (MIPS 24 $\mu$m) and \emph{Herschel} (PACS 70$\mu$m, 100$\mu$m, and 160$\mu$m). All ancillary data were convolved to $14^{\prime \prime}$ resolution and regridded to the same pixel scale (see \citealt{2018ApJ...860..165T} for more details). These processed infrared data allowed us to calculate the total infrared Luminosity $L_{\rm IR}$ (8-1000 $\mu$m) by combining the 24 $\mu$m, 70 $\mu$m, 100$\mu$m, and 160 $\mu$m luminosities \citep[]{2013MNRAS.431.1956G}.

\section{Results and Discussion}\label{sec:results and discussion}

\subsection{Correlation between Hot Gas and Dense Gas}\label{sec:dense-soft}

\begin{figure*}
    \centering 
    \vspace{0.9cm}
    \includegraphics[scale=0.905]{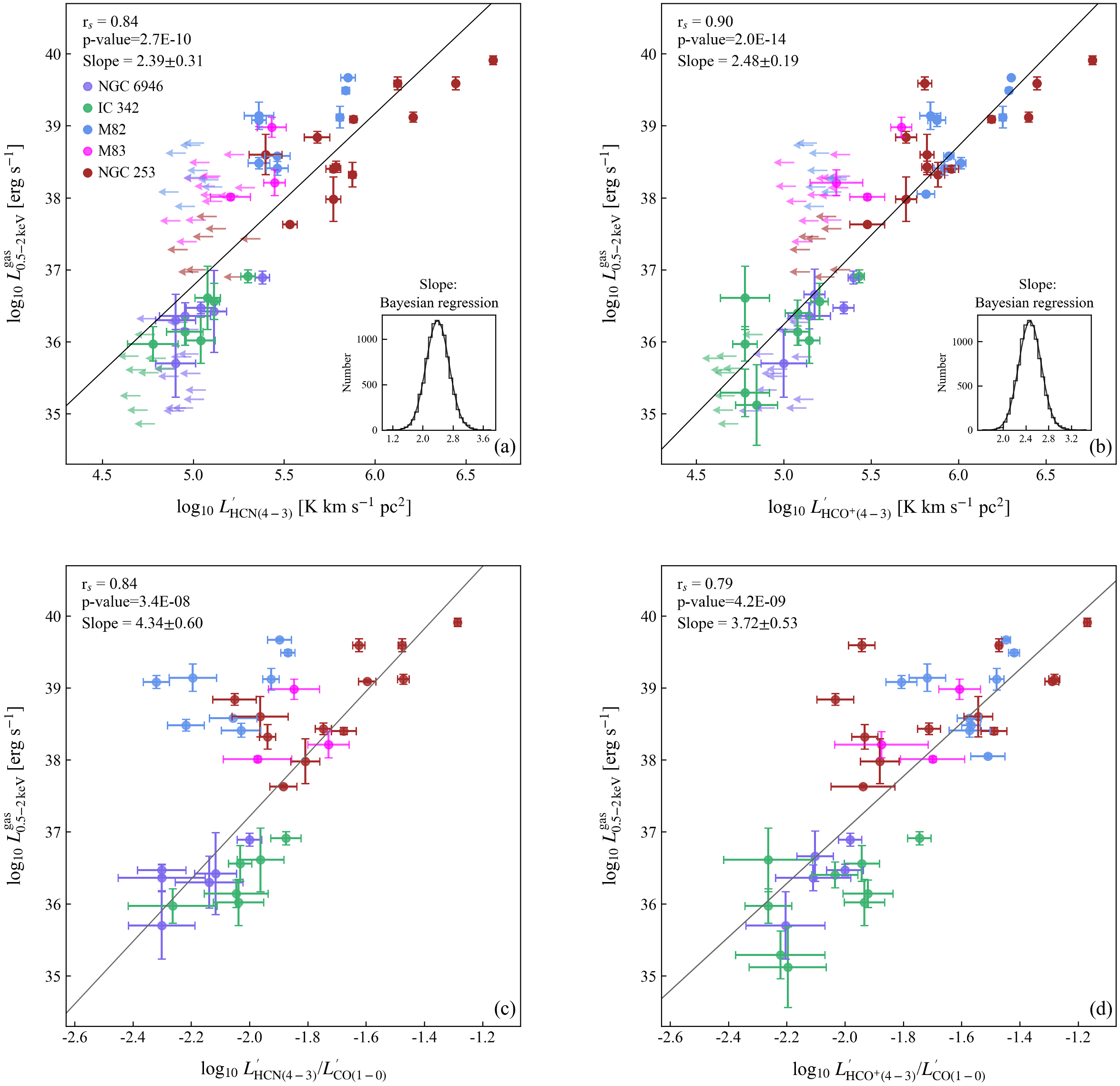}
    \vspace{0.05cm}
    \caption{Correlations between the diffuse X-ray emission (\lx) and the dense gas tracers. The colored circles represent the spatially resolved sub-kpc regions where HCN and HCO$^+$ are detected at S/N $\geqslant$ 3. {\it Top panels}: {\lx} as a function of $L'_{\rm HCN(4-3)}$ (left) and $L'_{\rm HCO^{+}(4-3)}$ (right), respectively. The solid lines in the panels indicate the best-fit relations derived from the Bayesian method for linear regression \citep{2007ApJ...665.1489K}. Arrows denote the 3$\sigma$ upper limits and are not included in the fitting. The inset shows the probability density distribution of the slope obtained by the Bayesian fitting. {\it Bottom panels}: Similar to the top panels, but the luminosities of HCN and HCO$^+$ are normalized by $L'_{\rm CO(1-0)}$, defined as $f_{\rm dense}$ (see Sect.~\ref{sec:dense-soft}). The 3$\sigma$ upper limits of $f_{\rm dense}$ are not displayed in these panels due to the relatively large scatter of the upper limits. The best-fit slope and the Spearman rank correlation coefficient for the {\lx} ${-}$ $L'_{\rm dense}$ and the {\lx} ${-}$ $f_{\rm dense}$ relation are listed in the top left of each panel, but note that the $f_{\rm dense}$ of M82 is ignored for {\hcnfour} due to its weak HCN emission.}
    \label{fig:1}
\end{figure*}

Figure~\ref{fig:1} shows the relationships between hot gas and dense gas on a sub-kpc scale ($\thicksim$ 230 pc) in the central region (within 2 kpc) of our five sample galaxies, covering a range of log{\lx} $\thicksim$ 5 dex and a range of log{\ldense} $\thicksim$ 2 dex. We adopt the Bayesian method code from the public IDL routine LINMIX\_ERR \citep{2007ApJ...665.1489K} for linear regression. In the code, the parameters were estimated by posterior median, and the error was adopted as the median absolute deviation of the posterior distribution. The density distributions of the fitted slopes are presented as insets in Figure~\ref{fig:1}. The best regression fits (not including the 3$\sigma$ upper limits) with uncertainties are listed below:
\begin{small}
\begin{equation}\label{hcn-lx}
    {\rm log}\,\frac{L_{\rm 0.5-2\,keV}^{\rm gas}} {\rm erg\,s^{-1}} = 2.39(\pm0.31)\,{\rm log}\,\frac{L'_{\rm HCN(4-3)}} {{\rm K\,km\,s^{-1}\,pc^{2}}} + 24.83(\pm1.72), 
    \end{equation}
    \end{small}
\begin{small}
\begin{equation}\label{hco-lx}
    {\rm log}\,\frac{L_{\rm 0.5-2\,keV}^{\rm gas}} {\rm erg\,s^{-1}} = 2.48(\pm0.19)\,{\rm log}\,\frac{L'_{\rm HCO^{+}(4-3)}} {{\rm K\,km\,s^{-1}\,pc^{2}}} + 23.84(\pm1.14).
    \end{equation}
    \end{small}

The solid lines in Figure~\ref{fig:1} (top panels) represent the fits for \hcnfour{} and \hcopfour{}, with Spearman correlation coefficients of $r_{\rm s}$=0.84 and $r_{\rm s}$=0.90, respectively. The corresponding $p$-values are $2.7\times10^{-10}$ and $2.0\times10^{-14}$. The results indicate that there is a strong connection between dense gas and hot gas in the nuclear regions of galaxies. This is consistent with our expectation because both dense gas and hot gas are closely interrelated to the star formation process \citep{2004gaoa,2004gaob,2005wujingwen,2012Mineo2,2014ApJ...784L..31Z,2015Usero,2018ApJ...860..165T}. 

\citet{2020MNRAS.494.1276J} found that the \hcnfour{} and \hcopfour{} emissions were concentrated at the inner $\thicksim$ 2 kpc of NGC 253, with a rapid drop until $\thicksim$ 0.5 kpc. \citet{2000Pietsch} found a similar concentration result that a third of diffuse soft X-ray emission of NGC 253 is in the nuclear area (1.1 kpc), which accounts for about 25\% of the total X-ray luminosity. Considering our Equation (\ref{hcn-lx}) or Equation (\ref{hco-lx}), these imply that the spatial distribution of hot gas in the central regions of galaxies may follow a similar trend to that of molecular gas, characterized by a sharp decrease at small radii and a slower decline further out.

Among the five sample galaxies, the HCN line luminosity of M82 is significantly lower than that of HCO$^+$, with a mean HCN/HCO$^+$ ratio of $\thicksim$ 0.3. We consider that this situation could be attributed to the low abundance of nitrogen in the sub-solar metallicity environment and/or the relatively low gas density condition of M82 \citep{origlia2004,nagao2011}. Recently, studies found the deficiency of nitrogen-bearing molecules was associated with the low metallicity in Large Magellanic Cloud and IC 10 \citep{Nishimura2016ApJ,Nishimura2016ApJ1}. Meanwhile, for low-metallicity Local Group galaxies, \citet{Braine2017} also observed weak HCN emission line compared to the {\hcop} emission. On the other hand, the lack of high density molecular gas \citep{1995Jackson} may also lead to the weak HCN emission, as {\hcop} is more easily collisionally excited to {$J=4 \rightarrow 3$} ($n_{\rm crit} \thicksim 1.3 {\times} {10^6} {\rm cm}^{-3}$) than HCN ($n_{\rm crit} \thicksim 5.6 {\times} {10^6} {\rm cm}^{-3}$ for {$J=4 \rightarrow 3$}) \citep{Schoier2005A&A}.

We define the ratios of \hcnfour{} / \coone{} and \hcopfour{} / \coone{} as proxies for the dense-gas fraction $f_{\rm dense}$ \citep{2020MNRAS.494.1276J}. The bottom panels of Figure~\ref{fig:1} show the diffuse X-ray luminosity as a function of $f_{\rm dense}$. A Spearman test yields a correlation coefficient of $r_{\rm s}$ = 0.84 with a $p$-value of $3.4\times10^{-8}$ for \hcnfour{}, and $r_{\rm s}$ = 0.79 with a $p$-value of $4.2\times10^{-9}$ for \hcopfour{}, suggesting a significant correlation between hot gas and dense-gas fraction. However, we ignore the $f_{\rm dense}$ of M82 for {\hcnfour} when analyzing the correlation between them, because the ratios of the weak HCN emission lines (discussed above) bring a huge dispersion and even lead to non-correlation. 

\subsection{Correlation between SFR and Hot Gas}\label{sec:hot gas and sfr}

\begin{figure}
    \includegraphics[scale=0.535]{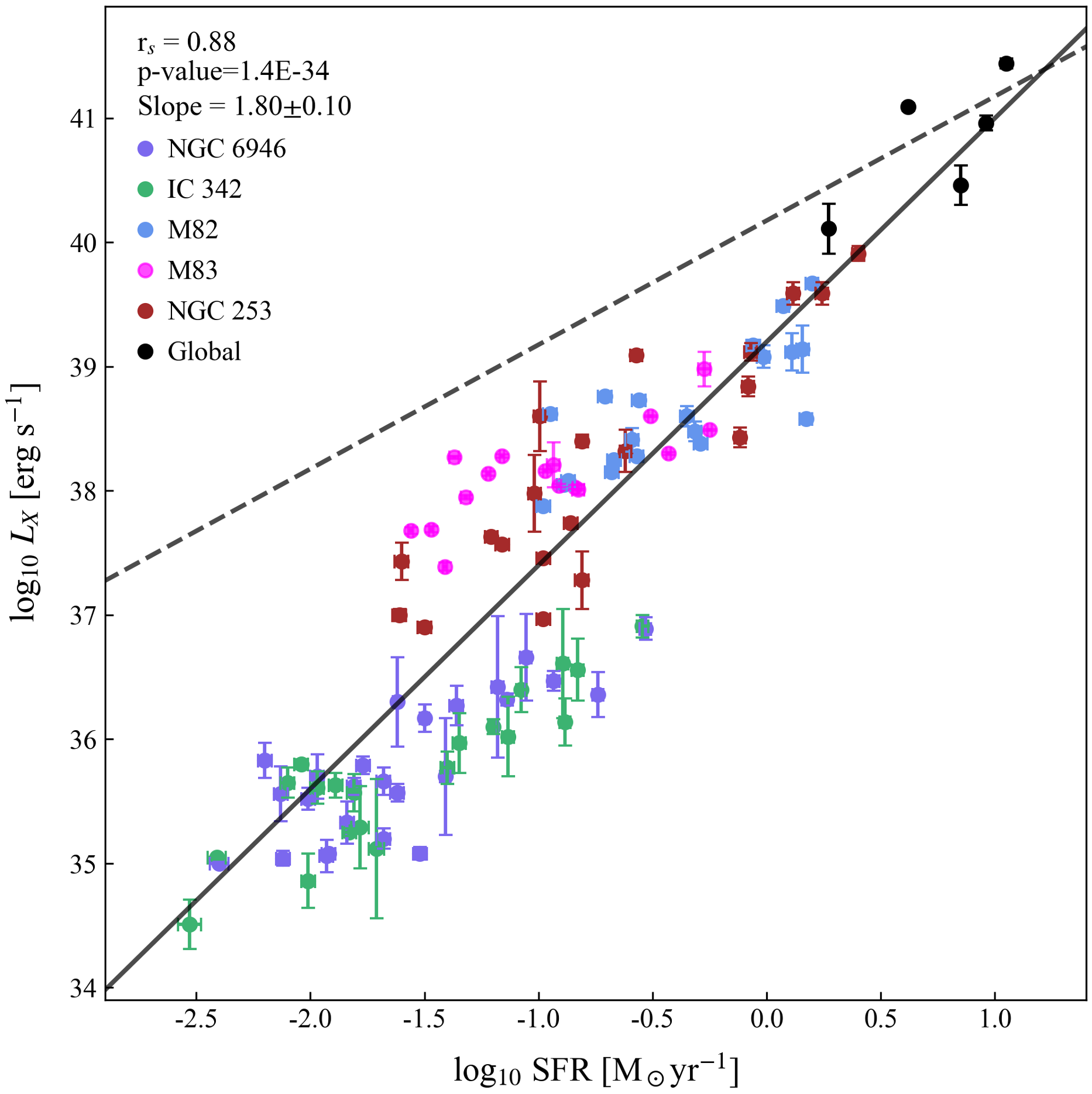}
    \caption{Correlations between X-ray luminosities and SFRs at the different spatial scales. The colored circles represent the spatially resolved sub-kpc positions in the central $50\arcsec \times 50\arcsec$ regions of our sample galaxies and the black circles represent the entire galaxies. The solid line indicates the best-fit of the {\lx} ${-}$ ${\rm SFR}$ relation at the sub-kpc scale (not including the black circles), with the fitting parameters listed in the top left corner of this plot. The dotted line shows the global linear relation between the intrinsic bolometric luminosity of hot gas and the SFR, as found by \citet{2012Mineo2}.} 
    \label{fig:2}
\end{figure}

Based on the calibrations of \citet{1998Kennicutt} and \citet{2011Murphy}, we use the total IR luminosity to calculate the SFR: 
\begin{equation}
    \left(\frac{{\rm SFR}}{M_\odot\ {\rm yr}^{-1}}\right) = 1.50\times 10^{-10} \left(\frac{L_{\rm IR}}{L_\odot}\right).
    \end{equation}
To investigate the correlation between hot gas and SFR on sub-kiloparsec scales, we analyze all JCMT beam matched positions (see Figure \ref{fig:f1a}) in the central $50\arcsec \times 50\arcsec$ regions of the five galaxies. The {\lx} $-$ SFR relation is shown in Figure~\ref{fig:2}, fitted using the Bayesian method code (see Sect.~\ref{sec:dense-soft}) which yields the following results:
\begin{small}
\begin{equation}
    {\rm log}\,\frac{L_{\rm 0.5-2\,keV}^{\rm gas}} {\rm erg\,s^{-1}} =1.80(\pm0.10)\,{\rm log}\,\frac{{\rm SFR}} {M_\odot\,{\rm yr}^{-1}} +39.16(\pm0.13),
    \end{equation}
    \end{small}
with a Spearman rank correlation coefficient of 0.88 and a $p$-value of $1.4\times10^{-34}$. This super-linear relation bears a larger dispersion of \emph{rms} = 0.69 dex than the dispersions (typically 0.3-0.4 dex) observed in the global-galaxy relations \citep{2012Mineo1,2012Mineo2,2014Mineo,2016lehmer}. \citet{2020Kouroumpatzakis} suggested that the scatter in their $L_{\rm X}-{\rm SFR}$ relation at sub-galactic scale ($\geq$ 1$\times$1 ${\rm kpc}^2$) could be the difference of the stellar population, while the local variations of stellar population and the associated X-ray emission could be smeared out when measured for the entire galaxy. For this study at higher resolution, the intrinsic scatter is more prominent, as more factors can have a large impact on our results in such a small area.

In the seminal work by \citet{2012Mineo2}, a global linear relation is found between the luminosity of unresolved diffuse emission and the SFR. For comparison, the relation, $L_{bol}$/SFR $\thicksim$ 1.5$\times$10$^{40}$ erg/s per $M_\odot$/yr, between the hot gas intrinsic bolometric luminosity and the SFR of them are shown in Figure~\ref{fig:2}. We also add the intrinsic bolometric luminosities of the five galaxies in Figure~\ref{fig:2} (black circles), and the detailed values are presented in Table \ref{tab:table1}. The intrinsic bolometric luminosities are derived from the 0.5 $-$ 2 keV luminosities using a bolometric correction factor of 2, as suggested by \citet{2012Mineo2}. It is clear that the global luminosities of the five galaxies are consistent with the linear relation, but there is a significant difference in our relations at low SFR. This difference implies that the hot gas environment may have dramatic variations from galactic center to the outskirts of the galaxy, thereby resulting in different relationships at different spatial scales. We consider there may be a large amount of hot gas concentrating in a small region of galaxies.

The contribution of several different types of unresolved compact sources (faint LMXBs, coronally active binaries (ABs), and cataclysmic variables (CVs)) to the total X-ray luminosity is relatively small and usually safely ignored in previous global studies. However, these unresolved stellar sources could become an important factor on small physical scales, especially for early type galaxies, where X-ray emission could be dominated by stellar sources \citep{Revnivtsev2007A&A}. Therefore, in order to determine the contribution of unresolved LMXBs and collective emission of ABs and CVs to the 0.5$-$2 keV band emission, we use \emph{K}-band luminosity to estimate the X-ray luminosity from these components based on the calibrations of \citet{2011Boroson}: 
\begin{small}
\begin{eqnarray}
    &L_{\mathrm{X}}(\mathrm{LMXBs})/L_{K}=10^{29.0 \, \pm \, 0.176} \, \mathrm{erg \, s^{-1}} L_{\mathrm{K\odot}}^{\, \, -1}, \\
    &L_{\mathrm{X}}(\mathrm{ABs+CVs})/L_{K}=4.4_{-0.9}^{+1.5} \times 10^{27} \, \mathrm{erg \, s^{-1}} L_{\mathrm{K\odot}}^{\, \, -1},
\end{eqnarray}
\end{small}
where $L_{\mathrm{K\odot}}$ is in solar luminosity. We find that, on average, the sum of the contribution of these faint compact souces accounts for about 20\% of the total soft X-ray luminosity, and the low contribution does not significantly change the {\lx} $-$ SFR relation. Therefore, we conclude that in the 0.5$-$2 keV band, the source-subtracted resolved emission is a reasonable estimate of the hot gas emission. Furthermore, we also find the contribution of ABs and CVs on sub-kiloparsec scales indeed dominates the unresolved X-ray emission (about twice as much as LMXBs), consistent with the results of M32 and the Galactic bulge reported by \citet{Revnivtsev2007A&A,Revnivtsev2007A&A_b,Revnivtsev2009Nature}. 

Last but not least, it is worth mentioning that all derived luminosities in Table~\ref{tab:table2} are determined in 2D. Projection effect is unavoidable in such measurements and may contribute to the scatters in Figure~\ref{fig:1} and~\ref{fig:2}. Although the inclination of galaxies in our samples ranges, the X-ray emission and the associated dense gas/star formation activity are physically linked, as we are working with the inner disk regions. In addition, as shown in Figure~\ref{fig:2}, the data appear to separate into two groups with a flatter slope but different intercepts. This may be due to the limited number of measurements. We will explore with future observations to examine these differences. If confirmed, it implies that the high SFR regions have a proportionally higher production efficiency for X-ray hot gas.

\section{Summary}\label{sec:summary}

We study the relationship between hot gas and star formation in inner $50\arcsec \times 50\arcsec$ region of five nearby star-forming galaxies as part of the JCMT program MALATANG with diffuse X-ray emission from hot gas using \emph{Chandra} archive data. All X-ray luminosities have been corrected for the Galactic and intrinsic absorption. We find significant correlations between $L'_{\rm dense}$ and {\lx} and between $f_{\rm dense}$ and {\lx}, with dense gas traced by both \hcnfour{} and \hcopfour{} at sub-kiloparsec scale. This indicates that the dense gas and the hot gas have a close relationship in the nuclear regions of galaxies. Combining with \emph{Spitzer} and \emph{Herschel} data, we also find a power-law relation for SFR and diffuse X-ray emission, with a super-linear slope of 1.80. However, the relation bears a rather large dispersion of \emph{rms} $\thicksim$ 0.69 dex. We attribute this to the enormous environment discrepancies at different regions of the galaxies. The complete MALATANG \uppercase\expandafter{\romannumeral2} survey as well as other large samples of galaxies observed at high resolution are indispensable to verifying these empirical relationships and to explain the physical connections behind them.


We thank the anonymous referee for providing constructive comments that improved our work. We acknowledge support by the National Key R\&D Program of China (Grant No. 2023YFA1607904) and the NSFC grants 12033004, 12333002, and 12221003. We are grateful to Dr. Jun-Zhi Wang, Zhi-Yu Zhang, Yang Gao, and Hao Chen for helpful discussions, and Dr. Xue-Jian Jiang for technical assistance in the data reduction. During this work, we were deeply saddened that Prof. Yu Gao passed away in 2022 May due to a sudden illness. C.Y.Z. will always remember Prof. Yu Gao's guidance and support. The East Asian Observatory operates the James Clerk Maxwell Telescope. 

This paper employs a list of \emph{Chandra} datasets, obtained by the \emph{Chandra} X-ray Observatory, contained in the \emph{Chandra} Data Collection (CDC) 230\dataset[doi:10.25574/cdc.230]{https://doi.org/10.25574/cdc.230}. We acknowledge the ORAC-DR, Starlink, CIAO, and GILDAS software for the data reduction and analysis. 


\bibliography{bibliography}{}
\bibliographystyle{aasjournal}

\end{document}